\documentclass{article}
\usepackage{spconf,amsmath,amsfonts,graphicx,booktabs}
\usepackage[dvipsnames]{xcolor}
\usepackage{diagbox}
\usepackage{multirow}

\usepackage[tableposition=top]{caption}
\usepackage{subcaption}
\usepackage{cite}

\usepackage{listings}
\usepackage{xcolor}

\usepackage{amssymb}
\usepackage{pifont}
\usepackage[hidelinks]{hyperref}
\newcommand{\cmark}{\ding{51}}%
\newcommand{\xmark}{\ding{55}}%

\definecolor{codegreen}{rgb}{0,0.6,0}
\definecolor{codegray}{rgb}{0.5,0.5,0.5}
\definecolor{codepurple}{rgb}{0.58,0,0.82}
\definecolor{backcolour}{rgb}{0.95,0.95,0.92}

\usepackage{enumitem}
\setlist[itemize]{leftmargin=*}
\setlist[enumerate]{leftmargin=*}

\usepackage{scrextend}

\title{GASS: generalizing audio source separation with large-scale data}

\name{Jordi Pons* \quad Xiaoyu Liu*\thanks{* Equal contribution} \quad Santiago Pascual \quad Joan Serr\`a}

\address{Dolby Laboratories}

\begin{document}
\ninept
\maketitle

\begin{abstract}

\vspace{1mm}

\noindent Universal source separation targets at separating the audio sources of an arbitrary mix, removing the constraint to operate on a specific domain like speech or music. Yet, the potential of universal source separation is limited because most existing works focus on mixes with predominantly sound events, and small training datasets also limit its potential for supervised learning. Here, we study a single general audio source separation (GASS) model trained to separate speech, music, and sound events in a supervised fashion with a large-scale dataset. We assess GASS models on a diverse set of tasks. Our strong in-distribution results show the feasibility of GASS models, and the competitive out-of-distribution performance in sound event and speech separation shows its generalization abilities. Yet, it is challenging for GASS models to generalize for separating out-of-distribution cinematic and music content. We also fine-tune GASS models on each dataset and consistently outperform the ones without pre-training. All fine-tuned models (except the music separation one) obtain state-of-the-art results in their respective benchmarks. 

\end{abstract}

\section{Introduction}
\label{sec:introduction}

Audio source separation consists of isolating the sources present in an audio mix. Most previous works frame the problem as a source-specific task, as in speech source separation~\cite{li2022efficient} (separating various speakers), or music source separation~\cite{rouard2023hybrid, luo2023music} (separating vocals, bass, and drums). 
For such tasks, a source-specific model is trained on dedicated datasets tailored to the task at hand. 
In contrast to source-specific separation tasks, universal source separation was recently proposed~\cite{kavalerov2019universal, wisdom2021s}, which consists of building source-agnostic models that are not constrained to a specific domain (like music or speech), and targets at separating an unknown number of sources given an arbitrary mix. However, existing universal source separation works predominantly focus on separating mixes similar to field recordings (with mostly sound events like dog barking or alarms).
Further, most supervised learning methods for this task rely on small training sets~\cite{tzinis2022compute, tzinis2020improving, kavalerov2019universal, postolache2023adversarial, wisdom2021s}. For instance, the commonly-used FUSS dataset contains only 23~hours of single-source recordings~\cite{wisdom2021s}. Considering the number of different sounds in the world, most audio sources might be under-represented in such small datasets. 
Hence, the potential of universal source separation is yet to be fully explored because (i) 
most previous works separate mixes with predominantly sound events instead of simultaneously separating a broader set of sources including speech, music, and sound events, 
and (ii) supervised universal source separation models have never been trained with large-scale data. 

Here, we explore training a unified model with large-scale data to address general audio source separation holistically\footnote{We use the term ``universal source separation'' when separating mixes with predominantly sound events (to be consistent with previous works~\cite{kavalerov2019universal}) and use the term ``general audio source separation'' when separating mixes containing speech, music, and/or sound events (our proposal).}, with the goal of separating any source from a given mix, including speech, music, and/or sound events. 
First, we scale up our audio source separation dataset by collecting 15,499~hours of recordings including speech, music, and sound events. 
Note that our dataset contains 3~orders of magnitude more data than FUSS~\cite{wisdom2021s}, the commonly-used dataset for supervised learning (Table~\ref{tab:numbers_data}). 
Next, to investigate the feasibility of general audio source separation$^1$, we train 3~state-of-the-art models with our large and diverse dataset. We are also interested in the generalization capabilities of the trained models. Hence, in addition to evaluating the models on different partitions of the same dataset (in-distribution), we also evaluate them on 4~standard downstream test sets, 
each one representing a different use case 
with different data and mixing pipelines (out-of-distribution). While in some cases the out-of-distribution results are competitive, in some others the separation results are not as satisfactory. Finally, we show that out-of-distribution performance can be improved by fine-tuning the pre-trained general audio source separation models on each task.

To our best knowledge, we offer the first study on supervised general audio source separation at scale without prior knowledge about the sources. Previous works~\cite{kong2023universal, liu2023separate} also consider speech and music in supervised universal source separation, 
but they assume the availability of a target source embedding to identify and separate the desired source from a mix.
Unsupervised approaches can also leverage large scale (noisy) data, but they tend to underperform supervised methods~\cite{wisdom2021sparse, dong2022clipsep, wisdom2020unsupervised}. Previous research also looked at the generalization capability of speech separation models~\cite{kadiouglu2020empirical, cosentino2020librimix}, but here we study the general audio source separation problem with a much more diverse set of out-of-distribution downstream tasks. Finally, our work is also conceptually similar to fine-tuning problem-agnostic self-supervised models~\cite{hsu2021hubert}, since we fine-tune source-agnostic audio separation models on source-specific tasks.

\section{Methodology}
\subsection{Creating a Large-scale Source Separation Dataset}
\label{sec:data_create}

We collect recordings from public and licensed datasets to scale up general audio source separation with $\approx$ 1.9 M recordings of speech, music, and sound events. 
We mix recordings $r_k$ at various gains $g_k$:
\begin{equation*}
	\label{eq:mixing}	
	m = \sum_{k=1}^{K} s_k = \sum_{k=1}^{K} g_k r_k ,
\end{equation*}
where we normalize $r_k$'s amplitudes to 1 before mixing, and $K$ is the number of resulting sources $s_k$ present in the mix. Note that $K$ is assumed to be unknown during training/inference and, following common practice~\cite{wisdom2021s}, we set $K\in\{1,2,3,4\}$. Also, defining what constitutes a source is a significant challenge. We find that the definition of ``any recording with one source'' might be impractical. For instance, considering separating two speakers talking in a cafeteria, it may be unnecessary to separate every individual sound in the background like the cutlery and the crowd noise. Similarly, in a mix with background music, it may not be desirable to separate out each instrument.
In our view, incorporating low-volume, non-dominant background sounds as a single, combined source to be separated together could enhance the realism of the resulting mixes. Hence, to build our dataset, we rely on the following definition: ``any recording with one source, except for low-volume background events that can contain one or more sources''. 
We distinguish between foreground and background sources by simply applying higher gains to foreground sources.
Table~\ref{tab:numbers_data} presents those gains $g_k$, together with the number of collected recordings $r_k$ and their source types:
\begin{itemize}
\item \textbf{Speech foreground} is a multilingual collection of public and licensed clean speech recordings, each with 1 speaker. A large portion of the recordings we use are public: AVSpeech~\cite{ephrat2018looking}, VCTK~\cite{veaux2017cstr}, DAPS~\cite{mysore2014can}, and TIMIT~\cite{garofolo1993timit}.

\item \textbf{Sound event foreground and background} are a combination of public and licensed datasets. The largest public dataset we use is (most of the content in) Freesound. Extensive listening finds that shorter Freesound recordings tend to be single-source, and longer ones tend to contain multiple sources. Hence, Freesound recordings shorter than 8\,sec are used as foreground, and longer ones as background.
We also use other background datasets, including: WHAM!~\cite{Wichern2019WHAMES} and DEMAND~\cite{thiemann2013diverse}.

\item \textbf{Music foreground and background} are a combination of public and licensed datasets. Public single-source datasets include: Slakh~\cite{manilow2019cutting}, ENST-drums~\cite{gillet2006enst}, VocalSet~\cite{wilkins2018vocalset}, QMUL singing database~\cite{black2014automatic}, MUSIC~\cite{zhao2018sop}, and EGFxSet~\cite{pedroza2022egfxset}. 
Hence, foreground music mostly contains vocals, bass, drums, guitar, and keys, but also includes synthesizers, percussion, and classical instruments. Background music includes licensed music mixes.
\end{itemize}
Note that our collection is significantly larger than FUSS~\cite{wisdom2021s}, the most common benchmark for universal source separation. 
After collecting our data, we define a set of rules to create the artificial mixes. These rules can be summarized into the following 3~upstream tasks:
\begin{itemize}
	\item \textbf{Speech separation}. These mixes always contain at least 1 speech foreground source. Other sources are sampled from the following sets: speech foreground, sound events foreground/background, and music background to create mixes for speech denoising and speech source separation (from 1 to 4 speakers) use cases. 
	
    \item \textbf{Sound event separation}. Sources are sampled from sound events foreground/background and music background to create mixes similar to previous universal source separation works~\cite{wisdom2021s,kavalerov2019universal}.
	
    \item \textbf{Music separation}. Sources are sampled from music foreground and sound events background to create mixes for music denoising and music source separation (from 1 to 4 sources) use cases.

\end{itemize}
Hence, to generate training data we randomly select: an upstream task (speech, sound event, or music separation with a probability of 0.25, 0.25, and 0.5, respectively), the number of sources $K$ (uniformly from 1 to 4), the recordings $r_k$ (which fragments and when they start in the mix), and the gains $g_k$ (sampled from a Beta distribution $Beta(2,1)$ within the ranges in Table~\ref{tab:numbers_data}). 
We then down-mix all the data to mono, zero-pad or truncate each sample to 8\,sec, and resample them to 48\,kHz.
Note that our large-scale dataset covers various sampling rates and bandwidths, all resampled to 48\,kHz, since 
we observe in preliminary experiments that models trained on this dataset perform competently at various (lower) sampling rates.

\begin{table}[t]
	\begin{center}
     \caption{Our large-scale general audio source separation dataset.}
      \vspace{-2ex}
		\resizebox{1.0\columnwidth}{!}{%
			\begin{tabular}{cccc}
                   \midrule
				Source type& $g_k$ (dB) & Single-source & \# Recordings  \\
				\midrule
				Speech foreground & [$-$10, 0 ] & \cmark & 759,397  \\
				
				Sound event foreground & [ $-$10, 0 ] & \cmark &   314,652 \\
				
				Sound event background & [ $-$20, $-$10 ] & \xmark &  398,360 \\
				
				Music foreground & [ $-$3, 0 ]  & \cmark &  ~~75,639 \\
				
				Music background & [ $-$20, $-$10 ]  & \xmark &  379,565 \\
				\midrule
				All dataset &  &  & 15,499 hours  \\
                FUSS~\cite{wisdom2021s} & & & 23 hours \\
                \midrule
			\end{tabular}
			
		}
		\label{tab:numbers_data}
	\end{center}
    \vspace{-6ex}
\end{table}

\label{sec:framework}

\subsection{Models and Upstream Training}
\label{sec:model}

\noindent \textbf{TDANet-Wav} (10.8\,M parameters). 
TDANet~\cite{li2022efficient} is a state-of-the-art waveform-based speech source separation model based on an encoder-separator-decoder architecture.
We adopt the official implementation\footnote{\url{https://github.com/JusperLee/TDANet}} and increase the encoder dimension to 1024 and proportionally double the dimension of the separator layers.

\vspace{1ex}

\noindent \textbf{TDANet-STFT} (7.4\,M parameters). We modify TDANet-Wav such that the encoder/decoder are replaced by STFT/iSTFT, and reuse the phase of the mixture for the iSTFT. The separator then outputs a mask over the STFT domain, not over a latent space as in TDANet-Wav.
We use 32 and 8\,ms frame length and stride, respectively. The bottleneck size is 
384 and the separator layers follow 
the recommended ratio of feature maps with respect to the bottleneck size~\cite{li2022efficient}.

\vspace{1ex}

\noindent \textbf{BSRNN} (21.8\,M parameters).
Band-Split RNN is a powerful model for music source separation~\cite{luo2023music} and speech enhancement~\cite{yu23b_interspeech}, also based on an encoder-separator-decoder architecture. Its encoder splits complex-valued STFT bins into bands and projects each band to a latent. We create 43 bands for our 48\,kHz model, 2 more bands on top of the setup proposed for separating vocals from music at 44.1\,kHz~\cite{luo2023music}. The separator consists of 12 interleaved band-level and sequence-level blocks with bidirectional LSTMs. The decoder undoes the band splitting and predicts complex-valued STFT masks.
We adopt an available open-source implementation\footnote{\url{https://github.com/sungwon23/BSRNN}}.

\vspace{1ex}
\noindent \textbf{IRM} (oracle). We compute the Ideal Ratio Mask (IRM) as an oracle upper bound using the magnitude STFT of the ground truth sources.

\vspace{1ex}

\noindent 
\textbf{Upstream training}. 
All models are trained on the upstream large-scale dataset for 10\,M steps using the Adam optimizer with a batch size of 10 and a cyclical learning rate 
between $10^{-7}$ and $10^{-4}$ spanning 400\,k steps per cycle.
All models predict 4 sources $\hat{s}_k$ given a mix $m$. When there are fewer targets during training ($K$$<$$4$), the extra targets are set to zeros. Permutation invariant training~\cite{kolbaek2017multitalker} (PIT) aligns the predictions with the targets, and we minimize the logarithmic-MSE loss with a threshold $\tau$ set to $-$30\,dB~\cite{wisdom2021s}: 
\begin{align*}	
		\mathcal{L}(s_k,\hat{s}_k) = \begin{cases}
			10 \log_{10}\left(\Vert \hat{{s}_k} \Vert^2 + \tau \Vert m \Vert^2\right) & \text{if } s_k = 0, \\ 10 \log_{10}\left(\Vert {s}_k - \hat{{s}}_k \Vert^2 + \tau \Vert {s}_k \Vert^2\right) & \text{otherwise.}
		\end{cases}
\end{align*}

\subsection{Evaluation Framework}
\label{sec:eval}

\noindent \textbf{Upstream (in-distribution) evaluation}.
For each upstream task (speech, sound event, and music separation), we set aside 3,000~mixes made of unseen recordings, which are sampled and mixed based on the same pipeline used for upstream training.

\vspace{1ex}

\noindent \textbf{Downstream (out-of-distribution) evaluation}.
We study the generalization capabilities of our models with out-of-distribution datasets. We consider the following 4 downstream tasks:

\begin{table}[t]
\centering
\caption{Upstream (in-distribution) results for speech, sound event, and music separation. SI-SDR column: SI-SDRs/SI-SDRi (dB). US/ES/OS: source count rate (\%).}
\vspace{-2ex}
\resizebox{1.0\columnwidth}{!}{
\begin{tabular}{cccccc}
    \toprule
    Task & Model & SI-SDR $\uparrow$ & US $\downarrow$ & ES $\uparrow$ & OS $\downarrow$\\
    \cmidrule{1-6}
    \multirow{ 3 }{*}[-3pt]{Speech} & TDANet-Wav & 53.5/\textbf{14.3} & \textbf{6.9} & \textbf{87.8} & 5.3 \\
    & TDANet-STFT & 80.6/13.8 & 14.1 & 83.3 & \textbf{2.6}\\
    & BSRNN & 44.3/12.8 & 13.8 & 80.1 & 6.1\\
    & IRM & 85.7/19.3 & 0 & 100 & 0\\
    \cmidrule{1-6}
    \multirow{ 3 }{*}[-3pt]{Sound events} & TDANet-Wav & 49.1/20.1 & 14.2 & 79.6 & 6.2 \\
    & TDANet-STFT & 71.8/\textbf{22.1} & 17.8 & 78.0 & \textbf{4.2}\\
    & BSRNN & 49.9/20.3 & \textbf{12.6} & \textbf{81.6} & 5.8\\
    & IRM & 78.0/28.3 & 0 & 100 & 0\\
    \cmidrule{1-6}
    \multirow{ 3 }{*}[-3pt]{Music} & TDANet-Wav & 52.6/14.6 & 5.9 & 90.9 & 3.2 \\
    & TDANet-STFT & 80.8/14.6 & 9.1 & 89.1 & \textbf{1.8}\\
    & BSRNN & 46.2/\textbf{18.2} & \textbf{3.9} & \textbf{93.2} & 2.9\\
    & IRM & 88.8/17.8 & 0 & 100 & 0\\
    \bottomrule
\end{tabular}
}
\label{tab:in_distribute}
\vspace{-3ex}
\end{table}

\begin{itemize}
	\item \textbf{FUSS} is a universal source separation dataset with 1 to 4 sources, with mixes at 16\,kHz similar to field recordings~\cite{wisdom2021s} (mostly sound events). We select the standard reverberated FUSS version for our downstream evaluation.
    Since FUSS is a subset of FSD50K~\cite{fonseca2021fsd50k}, we exclude FSD50K from our upstream dataset.
 	
    \item \textbf{Libri2Mix} is a common benchmark for speech source separation, with recordings at 16\,kHz containing 2 clean speech sources~\cite{cosentino2020librimix}. All LibriSpeech~\cite{librispeech} data is excluded from our upstream dataset.
	
    \item \textbf{DnR} dataset targets at separating cinematic mixes at 44.1\,kHz into speech, music, and sound effects~\cite{petermann2022cocktail}. Again, all involved datasets in DnR are excluded from our upstream dataset. Also note that DnR is a particular out-of-distribution case because it violates our source definition. We expect our models to separate each speaker, musical sources, and sound effect sources unless the music and sound effects are low-volume background events. However, DnR separates a mix into 3 combined stems: speech (with all speakers), music (with all musical sources), and sound effects (all together).

    \item \textbf{MUSDB} is a music source separation dataset at 44.1\,kHz with 4 sources: vocals, bass, drum, and `other'~\cite{musdb18-hq}. Yet, note that our models are trained to separate more musical sources, including vocals, bass, drums, keys, guitar, synthesizers, and classical instruments. Further, the `other' stem in MUSDB also violates our source definition, since such sources come grouped in one stem.
    We exclude both MUSDB and MedleyDB from our upstream data.
\end{itemize}
Although DnR (all stems) and MUSDB (`other' stem) violate our source definition, we are still interested in those to study fine-tuning a pre-trained (upstream) general model on a separation task defined differently. We conduct 3 evaluations for each downstream task:

\begin{itemize}
	\item \textbf{No-tuning}.~The pre-trained upstream models are assessed without any modification. This setup can also be seen as a zero-shot source separation case, where the models are pre-trained on a large dataset and then evaluated on new datasets without any adaptation.
 
	\item \textbf{Fine-tuning}.~The pre-trained upstream models are fine-tuned on the new downstream task alone with PIT. This setup studies the upstream model as a general model that can be pre-trained on a large dataset and then fine-tuned on a new use case. {When there are fewer training targets ($K$$<$$4$), the extra targets are set to zeros.}
 
	\item \textbf{From-scratch}.~The models are trained from-scratch on each downstream task.~This setup studies the performance of the models when they are not pre-trained on a large dataset.
\end{itemize}

\noindent Note, however, that the downstream datasets have different sampling rates. To unify our evaluation framework, we upsample the mixes and targets to 48\,kHz. In that way, we can compute the loss against the upsampled targets when fine-tuning and training from-scratch. To compute metrics with the original ground truth, we downsample the predicted sources back to the original sampling rates. In preliminary experiments, we observe that models trained from-scratch and evaluated in this way yield similar results as those obtained by models trained on the original datasets without resampling. 

\vspace{1ex}

\begin{table}[t]
\centering
\caption{Downstream (out-of-distribution) results on FUSS. SI-SDR column: SI-SDRs/SI-SDRi (dB). US/ES/OS: source count rate (\%).}
\vspace{-2ex}
\resizebox{1.0\columnwidth}{!}{
\begin{tabular}{cccccc}
    \toprule
    Evaluation & Model & SI-SDR $\uparrow$ & US $\downarrow$ & ES $\uparrow$ & OS $\downarrow$\\
    \cmidrule{1-6}
    \multirow{ 3 }{*}[-3pt]{No-tuning} & TDANet-Wav & 32.7/15.1 & 39.3 & 54.7 & \textbf{6.0} \\
    & TDANet-STFT & 30.0/\textbf{16.4} & 38.6 & 55.0 & 6.4\\
    & BSRNN & 30.5/16.0 & \textbf{36.6} & \textbf{57.0} & 6.4\\
    \cmidrule{1-6}
    \multirow{ 3 }{*}[-3pt]{Fine-tuning} & TDANet-Wav & 33.2/17.7 & \textbf{11.8} & 77.5 & 10.7 \\
    & TDANet-STFT & 34.0/18.1 & 16.5 & 73.1 & 10.4\\
    & BSRNN & 33.7/\textbf{18.6} & 14.0 & \textbf{78.5} & \textbf{7.5}\\
    \cmidrule{1-6}
    \multirow{ 3 }{*}[-3pt]{From-scratch} & TDANet-Wav & 33.0/13.7 & 22.2 & 65.2 & 12.5 \\
    & TDANet-STFT & 33.1/14.4 & 20.6 & 67.7 & \textbf{11.7}\\
    & BSRNN & 32.4/\textbf{14.4} & \textbf{13.7} & \textbf{70.6} & 15.7\\
    \cmidrule{1-6}
    SOTA & Postolache et al.~\cite{postolache2023adversarial} & 35.3/13.8 & 23.6 & 63.9 & 12.5 \\
    Oracle & IRM & 39.9/25.3 & 0 & 100 & 0\\
    \bottomrule
\end{tabular}
}
\label{tab:down_fuss}
\vspace{-3ex}
\end{table}

\noindent \textbf{Evaluation metrics}.
We use the standard metrics for each task:

\begin{itemize}
	\item \textbf{SI-SDR} (dB) in DnR. We use scale-invariant signal-to-distortion ratio~\cite{le2019sdr} (SI-SDR) to measure the quality of the separations.
 
	\item \textbf{SI-SDRs} (dB) in FUSS and upstream. For mixes with one source, we compute $\text{SI-SDRs} = \text{SI-SDR}(s_k, \hat{s}_k) = \text{SI-SDR}(m, \hat{s}_k)$~\cite{wisdom2021s}, since with one-source mixes the goal is to bypass the mix. The `s' sub-index stands for single-source.
 
	\item \textbf{SI-SDRi} (dB) in FUSS, Libri2Mix, and upstream. For mixes with 2 to 4 sources, we report $\text{SI-SDRi}=\text{SI-SDR}(s_k, \hat{s}_k)-\text{SI-SDR}(s_k, m)$~\cite{wisdom2021s, postolache2023adversarial}. The `i' sub-index stands for improvement. To account for inactive sources, estimate-target pairs that have silent target sources are discarded.
	 
    \item \textbf{US, ES, OS} (\%) in FUSS and upstream. Note that our models implicitly count the number of sources to separate. To evaluate source counting, we compute the proportion of the samples for which the number of nonzero predictions are fewer than (under-separation, US), equal to (equal-separation, ES), or more than (over-separation, OS) the number of nonzero targets~\cite{wisdom2021s}.
    A prediction is considered nonzero if its average energy is above $-$20\,dB relative to the softest nonzero target source~\cite{wisdom2021s}. 
	
    \item \textbf{SDR} (dB) in MUSDB. Defined in~\cite{stoter20182018}, is the per-source median across the median SDR over all 1\,second chunks in each song.
    
\end{itemize}

\section{Results}
\vspace{-1mm}
\noindent Separations produced by our models are available on our website\footnote{\url{http://www.jordipons.me/apps/GASS}}. 

\label{sec:exp}

\subsection{Upstream (In-distribution) Evaluation}
\label{sec:up_in}

Table~\ref{tab:in_distribute} lists the results for the 3~in-distribution tasks, showing that it is possible, with a single model, to perform general audio source separation (including speech, sound events, and music) without prior knowledge about the source types and the number of sources (up to 4). 
Comparing with the IRM, we see that the models are competitive. 
Interestingly, each model stands out at a different task: TDANet-Wav for speech separation, TDANet-STFT for sound event separation, and BSRNN for music separation. BSRNN outperforms IRM for music separation, showing the advantage of operating on the complex STFT for this task.
Also, the relatively high equal-separation rates (ES) show that the models are often able to count/separate the sources correctly. Among the miscounting cases, models tend to under-separate (US).
Finally, the high SI-SDRs values show that the models are able to bypass single-source inputs.

\begin{table}[t]
\vspace{-3ex}
\centering
\caption{Downstream (out-of-distribution) SI-SDRi (dB, $\uparrow$) results on Libri2Mix: speech source separation of 2 speakers.}
\vspace{-2ex}
\resizebox{0.98\columnwidth}{!}{
\begin{tabular}{lccc}
    \toprule
     & No-tuning & Fine-tuning & From-scratch \\
    \cmidrule{1-4}
    TDANet-Wav & \textbf{11.4} & \textbf{17.9} & \textbf{17.5} \\
    TDANet-STFT & 9.5 & 13.3 & 12.7 \\
    BSRNN & 8.7 & 16.0 & 15.2 \\
    \cmidrule{1-4}
    Li et al.~\cite{li2022efficient} (SOTA)& - & - & 17.4 \\
    IRM (Oracle) & 13.3 & - & - \\
    \cmidrule{1-4}    
    TDANet-Wav-FUSS & $-$6.6 & - & - \\    
    \bottomrule
\end{tabular}
}
\label{tab:down_libri}
\end{table}

\vspace{3mm}

\begin{table}[t]
\centering
\caption{Downstream (out-of-distribution) SI-SDR (dB, $\uparrow$) results on DnR for speech (S), music (M), and sound effects (FX).}
\vspace{-1ex}
\resizebox{1\columnwidth}{!}{
\begin{tabular}{l@{\hspace{1.5mm}}c@{\hspace{1mm}}c@{\hspace{1mm}}c@{\hspace{2mm}}c@{\hspace{1.5mm}}c@{\hspace{1.7mm}}c@{\hspace{1.5mm}}c@{\hspace{1.5mm}}c@{\hspace{1.5mm}}c}
     \multirow{2}{*}{} & \multicolumn{3}{c}{No-tuning} & \multicolumn{3}{c}{Fine-tuning} & \multicolumn{3}{c}{From-scratch} \\
    \cmidrule(lr){2-4}  \cmidrule(lr){5-7} \cmidrule(lr){8-10}
     & S & M & FX & S & M & FX & S & M & FX \\
     \midrule
     TDANet-Wav & \textbf{8.1} & \textbf{~~~0.6} & \textbf{$-$0.7} & \textbf{14.8} & 6.0 & 7.7 & \textbf{14.4} & 5.6 & 7.1 \\
     TDANet-STFT & 7.7 & $-$1.9 & $-$1.4 & 13.1 & 5.4 & 7.0 & 12.9 & 4.8 & 6.5 \\
     BSRNN & 7.9 & ~~~0.3 & $-$1.5 & 14.4 & \textbf{6.5} & \textbf{7.9} & 14.0 & \textbf{6.0} & \textbf{7.4} \\     
     \cmidrule{1-10}
     Unprocessed mixes & 1.0 & $-$6.8 & $-$5.0 & - & - & - & - & - & - \\
     Petermann et al.~\cite{petermann2022cocktail} & - & - & - & - & - & - & 12.3 & 4.2 & 5.7 \\
     IRM (Oracle) & 15.6 & ~~~8.5 & ~~10.7 & - & - & - & - & - & - \\
     \bottomrule  
\end{tabular}
}
\label{tab:down_dnr}
\vspace{-3ex}

\end{table}

\vspace{-4ex}
\subsection{Downstream (Out-of-distribution) Evaluation}
\label{sec:down_out}

\noindent \textbf{FUSS} (Table~\ref{tab:down_fuss}). First, the no-tuning SI-SDRi results consistently outperform those of the models trained from-scratch. This reflects that, for FUSS, the upstream models are capable to generalize. Yet, the under-separation rates are higher for the no-tuning models. We hypothesize that allowing low-volume multi-source backgrounds in the upstream tasks could cause under-separation in FUSS. After fine-tuning, however, we improve both the source counting accuracy and SI-SDRi, denoting how transferable to FUSS the upstream models are.
Note that the fine-tuned models are also significantly better than the state-of-the-art.

\vspace{1ex}

\noindent \textbf{Libri2Mix} (Table~\ref{tab:down_libri}). We observe that the best no-tuning model approaches the IRM result, denoting that the upstream model can generalize to the Libri2Mix task. Yet, there is still a gap when compared to the models trained from-scratch, which is not surprising considering the more general upstream task we address with the same model capacity.
Fine-tuning always outperforms training from-scratch, but the improvements are much smaller if compared to those 
obtained by fine-tuning on FUSS (Table~\ref{tab:down_fuss}). This shows that the no-tuning performance is indicative of how transferable the models are between tasks. Nonetheless, the fine-tuned TDANet-Wav obtains state-of-the-art results. 
We also evaluate a TDANet-Wav trained on FUSS (TDANet-Wav-FUSS) data to study the capacity of FUSS as an upstream dataset.
Its failure denotes the limitations of current supervised universal source separation to separate an arbitrary mix. 

\begin{table}[t]
\vspace{-3ex}
\centering
\caption{Downstream (out-of-distribution) SDR (dB, $\uparrow$) results on MUSDB for vocals (V), bass (B), drums (D), and other (O).}
\vspace{-2ex}
\resizebox{1.0\columnwidth}{!}{
\begin{tabular}{ccccccc}
    \toprule
    Evaluation & Model & V & B & D & O & Avg \\
    \cmidrule{1-7}
    \multirow{ 3 }{*}[-3pt]{No-tuning} & TDANet-Wav & \textbf{1.2} & 0.4 & 2.2 & 0.2 & 1.0 \\
    & TDANet-STFT & 0.7 & \textbf{1.1} & \textbf{3.2} & \textbf{0.3} & \textbf{1.3}\\
    & BSRNN & 0.0 & $-$0.2 & 0.0 & 0.0 & 0.0 \\
    \cmidrule{1-7}
    No-tuning & TDANet-Wav-M & 3.4 & 1.1 & 4.9 & 0.1 & 2.4 \\
    \cmidrule{1-7}
    \multirow{ 3 }{*}[-3pt]{Fine-tuning} & TDANet-Wav & 7.0 & \textbf{9.6} & \textbf{9.8} & 4.7 & \textbf{7.8} \\
    & TDANet-STFT & 6.8 & 5.9 & 6.5 & 4.4 & 5.9\\
    & BSRNN & \textbf{8.6} & 7.7 & 8.3 & \textbf{5.3} & 7.5 \\
    \cmidrule{1-7}
    \multirow{ 3 }{*}[-3pt]{From-scratch} & TDANet-Wav & 6.5 & \textbf{9.6} & \textbf{9.6} & 4.7 & \textbf{7.6} \\
    & TDANet-STFT & 6.7 & 6.1 & 6.4 & 4.1 & 5.8 \\
    & BSRNN & \textbf{8.2} & 7.4 & 8.4 & \textbf{5.2} & 7.3 \\
    \cmidrule{1-7}
    From-scratch & BSRNN w/ PIT  & 8.3 & 7.2 & 8.4 & 5.1 & 7.2 \\         
    \cmidrule{1-7}
    SOTA & Luo \& Yu~\cite{luo2023music}  & 10.0 & 7.2 & 9.0 & 6.7 & 8.2 \\ 
    Oracle & IRM & 9.4 & 7.1 & 8.5 & 7.9 & 8.2 \\
    \bottomrule
\end{tabular}
}
\label{tab:down_musdb}
\vspace{-4ex}
\end{table}

\vspace{1ex}
\noindent \textbf{DnR} (Table~\ref{tab:down_dnr}). The upstream models with no-tuning do not perform competently on this downstream task that
violates our source definition, since DnR aims at `3-group' separation but our models are trained to separate each source. For this reason, we observe high over-separation rates ($\approx95\%$ of the time, the 4th output contains non-negligible predictions).
When comparing no-tuning results with the unprocessed mixes, one notes that the no-tuning models are able to perform some degree of separation, but are much worse than the ones trained from-scratch. However, the fine-tuned models perform better than the ones trained from-scratch, indicating the transferability of the upstream models to a differently-defined task. {Note that we outperform Petermann et al.~\cite{petermann2022cocktail} (the best published result on DnR).}

\vspace{1ex}

\noindent \textbf{MUSDB} (Table~\ref{tab:down_musdb}).~On MUSDB, all no-tuning models perform poorly. First, we hypothesize that PIT may cause this problem, since we are not aware of prior works using PIT for music source separation. Hence, we compare two BSRNN models trained from-scratch with and without PIT\footnote{Our BSRNNs did not achieve the results as in Luo \& Yu~\cite{luo2023music} because we used a much smaller model and a single banding structure for all sources.}, to find out that their results are comparable and PIT is not the problem. Next, we train a TDANet-Wav with only upstream musical mixes to study the generalization capabilities of this music-specific model (TDANet-Wav-M). Yet, despite improving upon the general model, this model still performs much worse than the from-scratch TDANet-Wav. This fact, combined with the good in-distribution performance of our general models (Table~\ref{tab:in_distribute}), suggests that we have a mismatch between the upstream musical mixes and MUSDB mixes. Overall, these observations suggest future investigations, including collecting more music foreground data (note in Table~\ref{tab:numbers_data} that we only collected 75,639~music foreground recordings), increasing the model capacity, and probing interference between different upstream tasks.

\section{conclusion}
\label{sec:conclusion}
We studied general audio source separation models trained in a supervised fashion with large-scale data.~To study their generalization capabilities, we evaluated both in- and out-of-distribution performance. The in-distribution results show that the models are able to separate an unknown number of sources from a variate set of mixes that include speech, music, and sound events. Among the out-of-distribution results, the no-tuning models achieved competitive performance for sound event and speech separation, but we also noted that our models had challenges for generalizing to separate cinematic and music mixes. Moreover, with fine-tuning consistently outperforming from-scratch, we show how transferable the upstream models are to a diverse set of downstream tasks, even when there is a mismatch between the source definitions of the upstream and downstream tasks. All fine-tuned models (except the music separation one) obtain state-of-the-art results in their respective benchmarks.

\bibliographystyle{IEEEbib}
\bibliography{mybib}
\end{document}